\documentclass[10pt]{iopart}

\usepackage{graphicx}
\usepackage{dcolumn}
\usepackage{bm}

\usepackage{color}
\usepackage{siunitx}
\usepackage{hyperref}
\hypersetup{
	final,
	breaklinks,	
	pdftitle = {A two-photon photoemission study of \textit{Td}-WTe$_2$},
	pdfauthor = {Petra Hein},			
	linkcolor = black,							
	urlcolor = blue,						
	menucolor= red,							
	citecolor = blue,						
	filecolor = black,						
	bookmarksopen = true, 		
	bookmarksnumbered = true,	
	pdfpagemode = UseOutlines, 		
	pdfpagelayout = SinglePage,
	pdfmenubar = true,					
	pdftoolbar = true
}

\usepackage{geometry}

\newcommand{\overbar}[1]{\mkern 1.5mu\overline{\mkern-1.5mu#1\mkern-1.5mu}\mkern 1.5mu}

\makeatletter
\@namedef{ver@amsmath.sty}{}
\makeatother
\usepackage{amstext}
\usepackage{chemformula}
\usepackage[version=4]{mhchem}

\begin{document}
	
	\title[A combined laser-based ARPES and 2PPES study of \textit{Td}-WTe$_2$]{A combined laser-based ARPES and 2PPES study of \textit{Td}-WTe$_2$}
	
	\author{Petra Hein$^1$, Stephan Jauernik$^1$, Hermann Erk$^1$, Lexian Yang$^2$, Yanpeng Qi$^{3,4}$, Yan Sun$^4$, Claudia Felser$^4$ and Michael Bauer$^1$}
	
	\address{$^1$ Institute of Experimental and Applied Physics, University of Kiel, Leibnizstr. 19, D-24118 Kiel, Germany}
	\address{$^2$ State Key Laboratory of Low Dimensional Quantum Physics, Collaborative Innovation Center of Quantum Matter and Department of Physics, Tsinghua University, Beijing 100084, China}
	\address{$^3$ School of Physical Science and Technology, ShanghaiTech University, Shanghai 201210, China}
	\address{$^4$ Max Planck Institute for Chemical Physics of Solids, N\"othnitzer Str. 40, D-01187 Dresden, Germany}
	\ead{hein@physik.uni-kiel.de}
	
	\begin{abstract}
		Laser-based angle-resolved photoemission spectroscopy (ARPES) and two-photon photoemission spectroscopy (2PPES) are employed to study the valence electronic structure of the Weyl semimetal candidate \textit{Td}-WTe$_2$ along two high symmetry directions and for binding energies between $\approx\SI{-1}{\electronvolt}$ and \SI{5}{\electronvolt}. The experimental data show a good agreement with band structure calculations. Polarization dependent measurements provide furthermore information on initial and intermediate state symmetry properties with respect to the mirror plane of the \textit{Td} structure of WTe$_2$.  
	\end{abstract}
	
	
	\submitto{\JPCM}
	\maketitle
	\ioptwocol
	
	\section{\label{sec:level1}Introduction}
	
\begin{figure}
		\centering
		\includegraphics[width=1\linewidth]{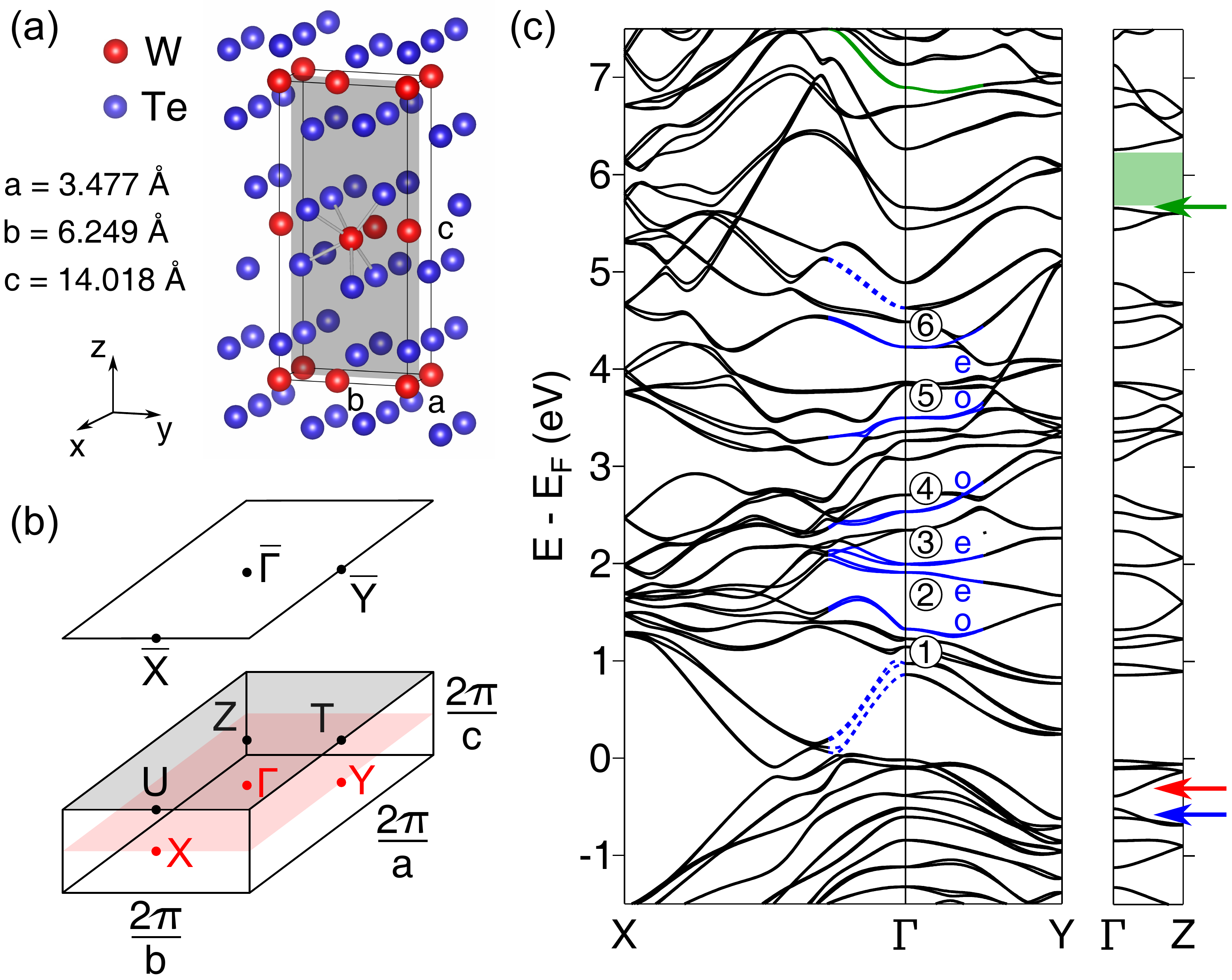}
		\caption{Crystal and electronic structure of \textit{Td}-WTe$_2$. (a) Orthorhombic unit cell with experimental lattice constants at $\text{T} = \SI{113}{\kelvin}$ \cite{Mar1992}. The mirror plane (\textit{b}-\textit{c} plane) of the \textit{Td} structure is indicated. (b) 3D Brillouin zone and its projection onto the (001) surface. High symmetry planes $\Gamma\text{XY}$ and $\text{ZUT}$ are indicated in red and gray, respectively. (c) Calculated bulk band structure along selected high symmetry directions. The bands highlighted by blue dashed lines are observed in the bichromatic 2PPES spectra. The bands highlighted by blue and green solid lines are observed in the monochromatic 2PPES spectra. Even (e) and odd (o) symmetry of these bands with respect to the mirror plane of the \textit{Td} structure of WTe$_2$ is indicated. The numbering of the bands is introduced for comparison with the experimental results shown in Fig. \ref{fig:figure_05} and Fig. \ref{fig:figure_06}. Details regarding the three markers and the shaded area in the band structure data along  $\Gamma$-$\text{Z}$ are given in the text.}
		\label{fig:figure_01}
	\end{figure}

	As a potential type-II Weyl semimetal~\cite{Soluyanov2015}, the transition metal dichalcogenide WTe$_2$ attracted considerable interest in recent years. Even though characteristic Weyl points and Fermi arcs in the electronic structure could not directly be observed so far~\cite{Bruno2016,Wang2016,Sanchez2016,Feng2016,Wu2016,DiSante2017,Zhang2017}, 
	several experimental studies clearly evidenced the topologically nontrivial phase in this material \cite{Li2017, Lin2017, Lv2017, Li2019}. WTe$_2$ crystallizes in a distorted 1\textit{T} structure (\textit{Td} structure) with an orthorhombic unit cell as shown in Fig. \ref{fig:figure_01}(a) \cite{Brown1966}. It consists of covalently bonded Te-W-Te trilayers that are held together via weak van der Waals forces resulting in a natural cleavage plane perpendicular to the layers in [001] direction. The bulk Brillouin zone (BZ) and its (001) surface projection are shown in Fig. \ref{fig:figure_01}(b). Results of a band structure calculation along three selected high symmetry directions of the bulk Brillouin zone are shown in Fig. \ref{fig:figure_01}(c). Further band structure data are provided in Fig. S1 of the supplemental material.  
	
	In this work, we report on a combined laser-based ARPES and 2PPES study of the occupied and unoccupied electronic structure of \textit{Td}-WTe$_2$. The combination of these complementary photoemission techniques enables us to access a binding energy range between $\approx$ \SI{-1}{\electronvolt} and \SI{5}{\electronvolt} with respect to the Fermi energy $E_\text{F}$. Analysis of the polarization dependence of ARPES and 2PPES spectra reveal information on the symmetry of the probed occupied and unoccupied states with respect to the mirror plane of the \textit{Td} structure. The data additionally provide insights into the final state electronic structure above the vacuum energy $E_\text{V}$. Overall, band energies, band dispersions, and symmetry properties as determined from the experimental data show a good agreement with orbital-resolved band structure calculations. 

	\section{\label{sec:level2}Experimental details and theoretical method}
	
	High-quality \textit{Td}-WTe$_2$ crystals were grown using a chemical vapour transport technique. Stoichiometric tungsten powder (\SI{99.9}{\percent}) and tellurium powder (\SI{99.99}{\percent}) were ground together and loaded into a quartz tube with a small amount of the transport agent TeBr$_4$. All weighing and mixing was carried out in a glove box. The tube was sealed under vacuum and placed in a two-zone furnace. The hot zone and the cold zone were maintained for one week at a constant temperature of \SI{800}{\celsius} and \SI{700}{\celsius}, respectively.
	Prior to the photoemission experiments, the \textit{Td}-WTe$_2$ samples were cleaved at room temperature under ultrahigh vacuum (UHV) conditions (\SI{3e-10}{\milli\bar}). From the low-energy cutoff of the ARPES spectra, we determined a value of $\Phi\approx\SI{4.9}{\electronvolt}$ for the work function of the samples.
	
\begin{figure}[t]
		\centering
		\includegraphics[width=1\linewidth]{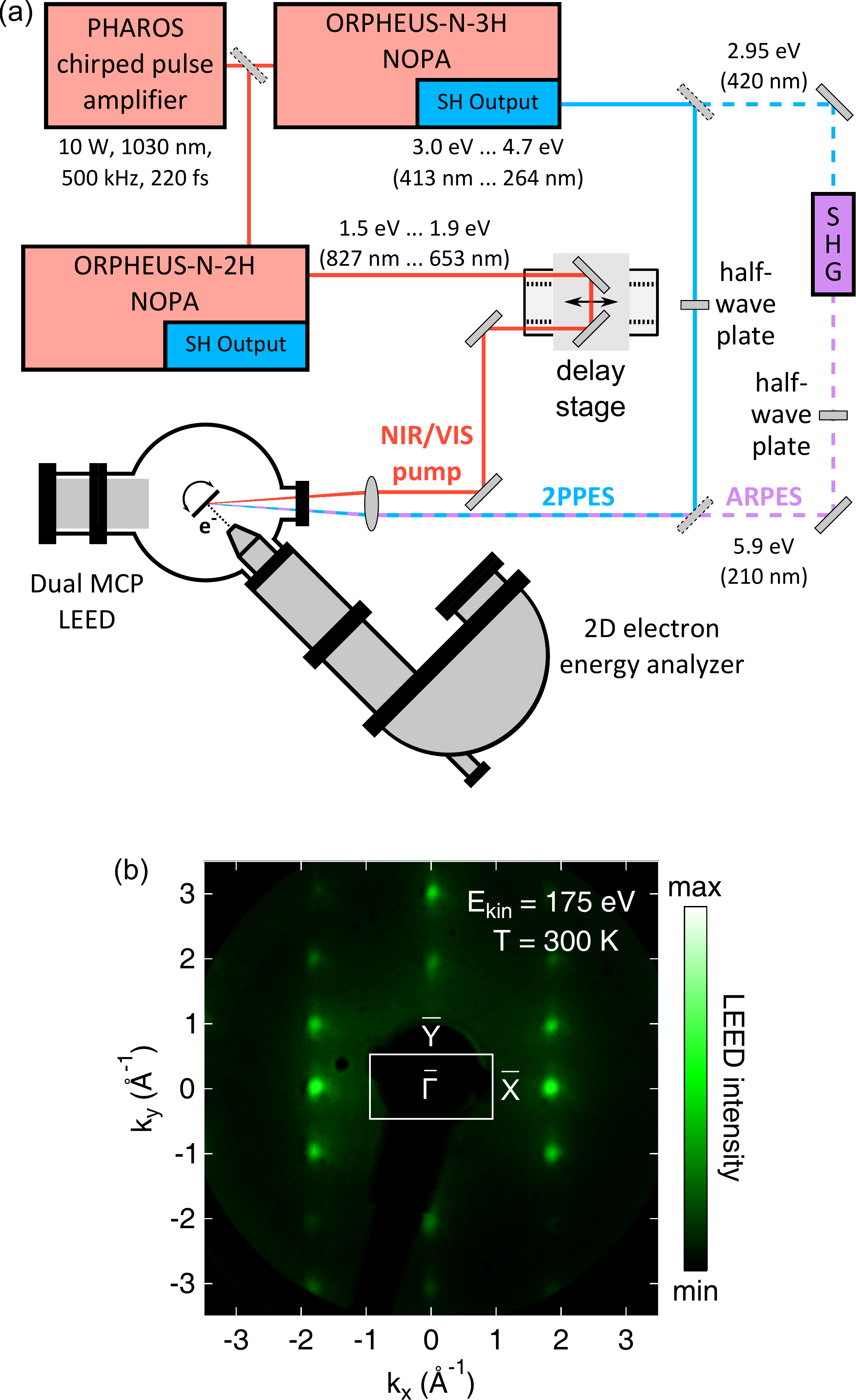}
		\caption{(a) Scheme of the combined laser-based ARPES and 2PPES setup. Details are described in the text. (b) LEED image of the (001) surface of \textit{Td}-WTe$_2$ with the surface Brillouin zone indicated.}
		\label{fig:figure_02}
	\end{figure}

	A scheme of the experimental setup, consisting of a tunable femtosecond laser system and an UHV chamber, is shown in Fig. \ref{fig:figure_02}(a) \cite{Jauernik2018}. ARPES and 2PPES measurements were performed with the second harmonic output of a non-collinear optical parametric amplifier (NOPA) (ORPHEUS-N-3H, Light Conversion) that is pumped by the third harmonic of a chirped pulse amplifier (PHAROS, Light Conversion). For monochromatic 2PPES experiments, the photon energy $h\nu$ was varied between \SI{3.0}{\electronvolt} (\SI{413}{\nano\meter}) and \SI{4.7}{\electronvolt} (\SI{264}{\nano\meter}) in steps of \SI{0.1}{\electronvolt}. For the ARPES experiments, we used \SI{5.9}{\electronvolt} (\SI{210}{\nano\meter}) light pulses that were produced by second harmonic generation (SHG) of the \SI{2.95}{\electronvolt} (\SI{420}{\nano\meter}) output of the NOPA system. The \SI{5.9}{\electronvolt} pulses were also used as probe pulses in bichromatic 2PPES experiments. \SI{30}{\femto\second} near infrared (NIR) / visible (VIS) pump pulses were generated by an additional NOPA (ORPHEUS-N-2H, Light Conversion) that is pumped by the second harmonic of the chirped pulse amplifier. Pump photon energies were varied between \SI{1.5}{\electronvolt} (\SI{827}{{\nano\meter}}) and \SI{1.9}{\electronvolt} (\SI{653}{{\nano\meter}}). The maximum incident pump fluence on the sample surface was $\approx \SI{110}{\micro\joule\per\square\centi\meter}$.
	The polarization of the laser light could be switched from s- to p-polarization using tunable zero-order half-wave plates. Here, s- (p-) polarization denotes the case that the electric  field  vector lies perpendicular to the plane of incidence (in the plane of incidence) held by the sample surface normal and the direction of incidence of the laser light.  
	
	Photoelectron spectra were recorded with a SPECS Phoibos 100 analyzer. For the ARPES and bichromatic 2PPES experiments, the total energy resolution was \SI{40}{\milli\electronvolt}. For the monochromatic 2PPES experiments, the total energy resolution was \SI{74}{\milli\electronvolt} to \SI{90}{\milli\electronvolt}, depending on the photon energy. 
	The quality of sample cleavage and the crystallographic orientation of the sample surface was checked by low-energy electron diffraction (LEED) (BDL600IR MCP2, OCI Vacuum). For the photoemission experiments, the \textit{Td}-WTe$_2$ samples were aligned either in $\overbar{\Gamma}$-$\overbar{\text{X}}$ or in $\overbar{\Gamma}$-$\overbar{\text{Y}}$ direction of the (001) surface Brillouin zone with respect to the analyzer entrance slit [see LEED image in Fig. \ref{fig:figure_02}(b)].  
	All measurements were performed at room temperature.

	The electronic band structure was calculated by \textit{ab-initio} calculation based on density functional theory (DFT) with Projector augmented-wave (PAW) method \cite{Bloechl1994} as implemented in the Vienna \textit{Ab-initio} Simulation Package (VASP) \cite{Kresse1996}. The exchange and correlation energies were considered on the level of the generalized gradient approximation (GGA) with a Perdew-Burke-Ernzerhof (PBE) functional \cite{Perdew1996}. The energy cutoff was set to be \SI{350}{\electronvolt} for the plane wave basis. The experimental lattice constants from Ref. \cite{Mar1992} were used in all the calculations [see Fig. \ref{fig:figure_01}(a)].
	
	\section{\label{sec:level3}Results and discussion}
	
	\begin{figure*}
		\centering
		\includegraphics[width=0.74\linewidth]{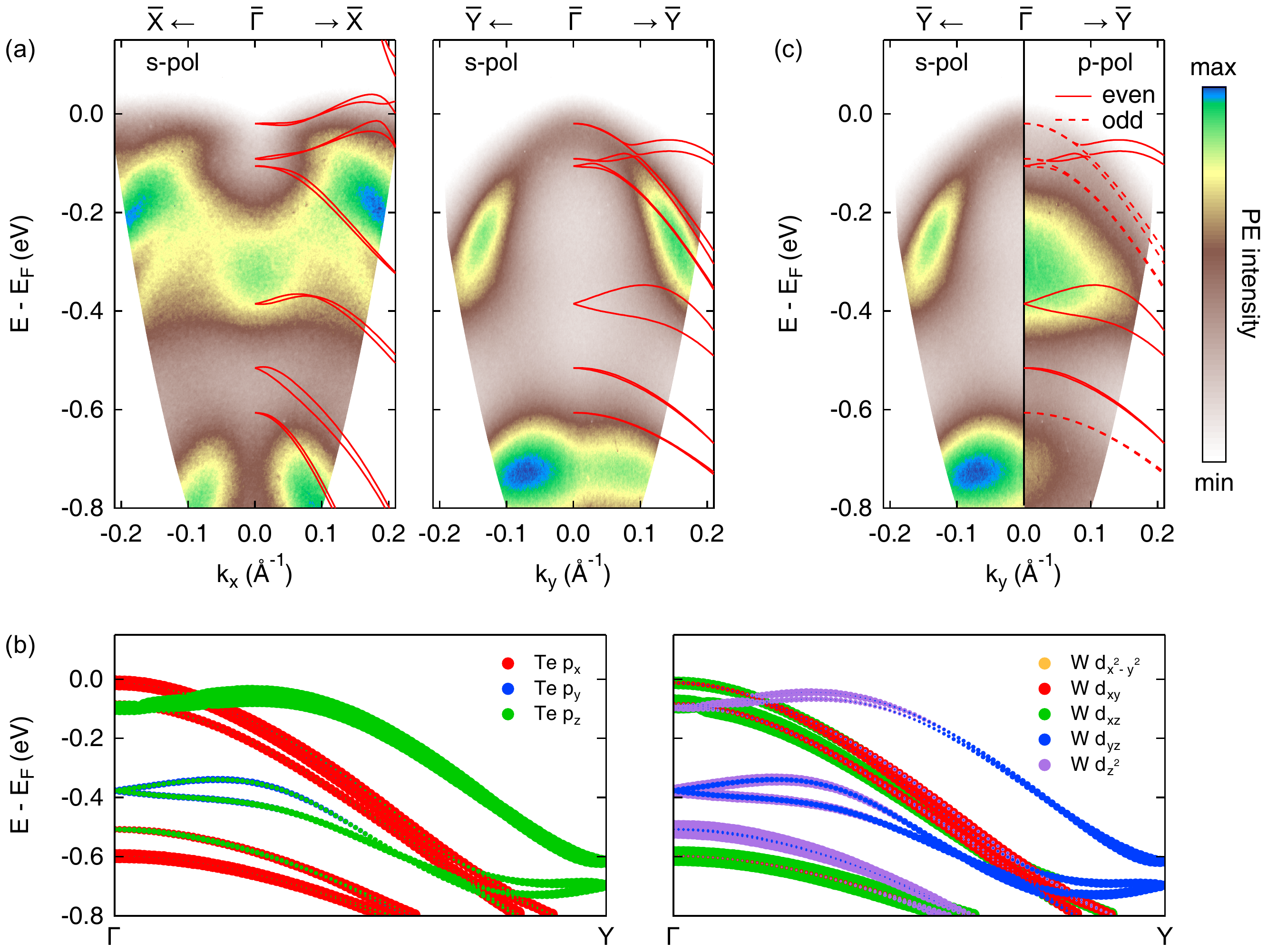}
		\caption{Laser-based ARPES data of \textit{Td}-WTe$_2$ in comparison to band structure calculations. (a) ARPES intensity maps (binding energy $E-E_{\text{F}}$ vs. surface parallel momentum $k_{\text{x}}$ and $k_{\text{y}}$, respectively) along $\overbar{\Gamma}$-$\overbar{\text{X}}$ and $\overbar{\Gamma}$-$\overbar{\text{Y}}$ in comparison to calculated band structures along $\Gamma$-$\text{X}$ and $\Gamma$-$\text{Y}$. Spectra were recorded with s-polarized light at $h\nu=$\SI{5.9}{\electronvolt}. (b) Calculated Tellurium and Tungsten contributions to the bands along $\Gamma$-$\text{Y}$ in the energy range covered by the ARPES data. The size of the data points indicates the respective orbital contribution to the individual bands. In Fig. S2 of the supplemental material, the orbital-resolved data are shown separately. (c) Comparison of ARPES intensity maps along $\overbar{\Gamma}$-$\overbar{\text{Y}}$ recorded with s-polarized light (left) and p-polarized light (right) at $h\nu=$\SI{5.9}{\electronvolt}. Results of the band structure calculations along $\Gamma$-$\text{Y}$ are added with solid lines indicating bands of even symmetry and dashed lines indicating bands of odd symmetry with respect to the mirror plane of the \textit{Td} structure.}
		\label{fig:figure_03}
	\end{figure*}
	
	\subsection{ARPES results}
	
	Figure \ref{fig:figure_03}(a) shows ARPES intensity maps of \textit{Td}-WTe$_2$ along $\overbar{\Gamma}$-$\overbar{\text{X}}$ and $\overbar{\Gamma}$-$\overbar{\text{Y}}$ recorded using s-polarized light. Overall, the spectra conform with results of other ARPES studies that used similar photon energies of $h\nu \approx \SI{6}{\electronvolt}$~\cite{Bruno2016, Wu2016, Caputo2018}.
	For comparison, the experimental data are overlaid with results of the band structure calculations along $\Gamma$-$\text{X}$ and $\Gamma$-$\text{Y}$. 
	Band energies and band dispersions predicted by the calculations are reproduced reasonably well. Due to matrix element effects, some of the bands are not visible in the spectra \cite{Wang2016}.
	As expected for a measurement in s-polarization geometry and within the free electron final state approximation, we observe for most bands a considerable reduction in the photoemission yield at $\overbar{\Gamma}$, i.e., for normal emission \cite{Moser2017}. However, in the binding energy range between \SI{-0.2}{\electronvolt} and \SI{-0.4}{\electronvolt} along $\overbar{\Gamma}$-$\overbar{\text{X}}$, the signal becomes maximum for normal emission. We suspect final state effects being responsible for this observation: The results of the band structure calculations indicate that we observe here a resonant transition to bulk final states at $E-E_\text{F} \approx \SI{5.6}{\electronvolt}$ indicated by the green marker in Fig. \ref{fig:figure_01}(c). These states show only a weak dispersion in normal direction, i.e., along $\Gamma$-$\text{Z}$, implying that this transition is not well described by the free electron final state approximation.
	
	\begin{figure*}[h]
		\centering
		\includegraphics[width=1\linewidth]{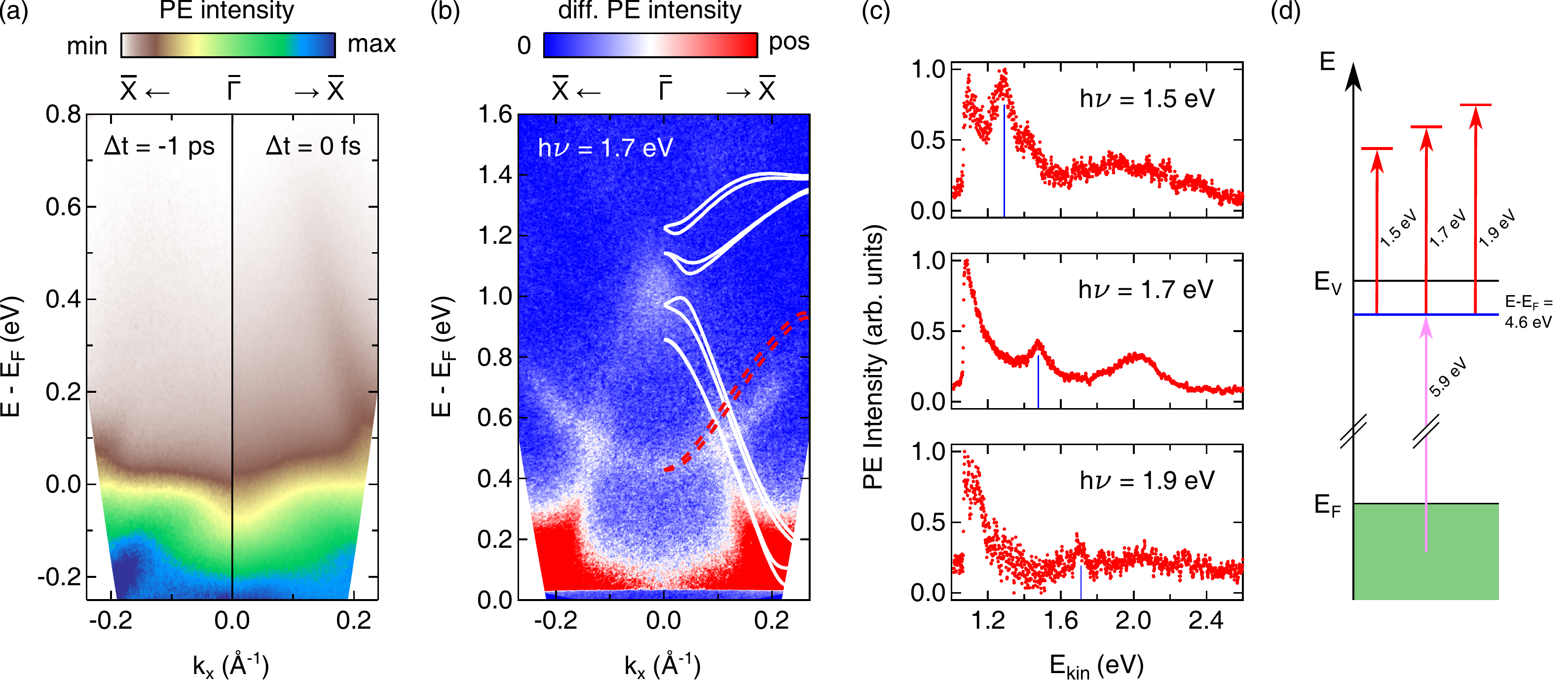}
		\caption{Bichromatic 2PPES data of \textit{Td}-WTe$_2$. (a) Comparison of bichromatic 2PPES intensity maps (binding energy $E-E_{\text{F}}$ vs. surface parallel momentum $k_\text{x}$) along $\overbar{\Gamma}$-$\overbar{\text{X}}$ recorded before the excitation with the pump pulse ($\Delta t = \SI{-1}{\pico\second}$) and at temporal overlap between pump and probe pulse ($\Delta t = \SI{0}{\femto\second}$). The data were recorded with p-polarized pump pulses ($h\nu = \SI{1.7}{\electronvolt}$) and s-polarized probe pulses ($h\nu = \SI{5.9}{\electronvolt}$). (b) Difference 2PPES intensity map along $\overbar{\Gamma}$-$\overbar{\text{X}}$ compiled from subtraction of data recorded at $\Delta t = \SI{-1}{\pico\second}$ and $\Delta t = \SI{0}{\femto\second}$. The data cover an excited energy range up to a binding energy of $E-E_\text{F}=\SI{1.6}{\electronvolt}$. Results of band structure calculations along $\Gamma$-$\text{X}$ are included as white lines. The red dashed line shows results of the band structure calculations along $\Gamma$-$\text{X}$ at $E-E_\text{F}\approx \SI{4.6}{\electronvolt}$ for comparison with the electron-like parabola visible in the difference 2PPES intensity map. Note the different energy scales in (a) and (b). (c) Difference EDCs deduced from difference 2PPES intensity maps for different pump photon energies. The blue markers indicate the shift in energy of the minimum of the electron-like parabola. 2PPES intensities were integrated over a momentum window of $\SI{\pm 0.025}{\per\angstrom}$ around $\overbar{\Gamma}$. (d) Schematic illustration of the 2PPE process with the NIR / VIS pulses acting as probe pulses.
		} 
		\label{fig:figure_04}
	\end{figure*} 
	
	The ARPES data along the $\overbar{\Gamma}$-$\overbar{\text{Y}}$ direction was taken with the UV light incident in the mirror plane of the \textit{Td} structure of WTe$_2$ so that the analysis of the polarization dependence allows for conclusions on the symmetry of the probed initial states with respect to the mirror plane \cite{Huefner1995, Mulazzi2006}. ARPES spectra recorded with p-polarized (s-polarized) light probe initial state wave functions with even (odd) symmetry. The comparison with orbital-resolved band structure data shown in Fig. \ref{fig:figure_03}(b) enables us to identify the orbital contributions probed with the different polarizations: $d_{yz}$, $d_{x^2-y^2}$, $d_{z^2}$, $p_y$, and $p_z$ are even with respect to reflection in the mirror plane of the \textit{Td} structure, whereas $d_{xy}$,  $d_{xz}$, and $p_x$ orbitals are odd with respect to reflection in the mirror plane of the \textit{Td} structure. Fig. \ref{fig:figure_03}(c) directly compares ARPES intensity maps along $\overbar{\Gamma}$-$\overbar{\text{Y}}$ recorded with s-polarized light and p-polarized light, respectively. The distinct differences in the spectra result to large extent from the dipole selection rules. The parabolic band observed close to $E_\text{F}$ with s-polarized light conforms with the predominant $\text{Te}$-$p_x$ and $\text{W}$-$d_{xz}$ character of the hole-like bands in this energy range [see Fig. \ref{fig:figure_03}(b)]. The second band visible in the s-polarized data can be assigned to the hole-like band at a binding energy of $E_B \approx \SI{-0.6}{\electronvolt}$, which also shows a predominant $\text{Te}$-$p_x$ and $\text{W}$-$d_{xz}$ character. The weak photoemission signal observed in the p-polarized data in this energy range conforms with the $\text{Te}$-$p_z$ and $\text{W}$-$d_{z^2}$ orbital character of the next higher lying band along $\Gamma$-$\text{Y}$. Note that both bands are downward dispersing along $\Gamma$-$\text{Z}$ and cross each other halfway between $\Gamma$ and $\text{Z}$ [see blue marker in Fig. \ref{fig:figure_01}(c)]. In general, an ARPES experiment probes a momentum cut at a finite value of the wave vector perpendicular to the surface projected BZ that depends on photon energy, crystal inner potential, electron momentum parallel to the surface, and final state effective mass \cite{Huefner1995}. The appearance of the two bands at similar binding energies and shifted to higher binding energies in comparison to the results of the band structure calculations along $\Gamma$-$\text{Y}$ may therefore indicate that in our experiment we probe the band structure at a finite momentum value in the surface normal direction. The data recorded with p-polarized light is finally dominated by a broad spectral feature extending from $E-E_{\text{F}} \approx$ \SI{-0.2}{\electronvolt} to \SI{-0.4}{\electronvolt}. The band structure calculations predict in this energy range two bands with even symmetry formed from $\text{Te}$-$p_z$, $\text{W}$-$d_{z^2}$, and $\text{W}$-$d_{yz}$ orbitals, which merge at $\Gamma$. These bands exhibit a strong dispersion along $\Gamma$-$\text{Z}$ [see red marker in Fig. \ref{fig:figure_01}(c)] covering an overall energy range of $\approx \SI{0.25}{\electronvolt}$. We assume that the observed spectral broadening results from a combination of the $\Gamma$-$\text{Z}$ dispersion and a limited momentum resolution of the ARPES experiment normal to the sample surface.

	\subsection{Bichromatic 2PPES results}
	
	\begin{figure*}[]
		\centering
		\includegraphics[width=1\linewidth]{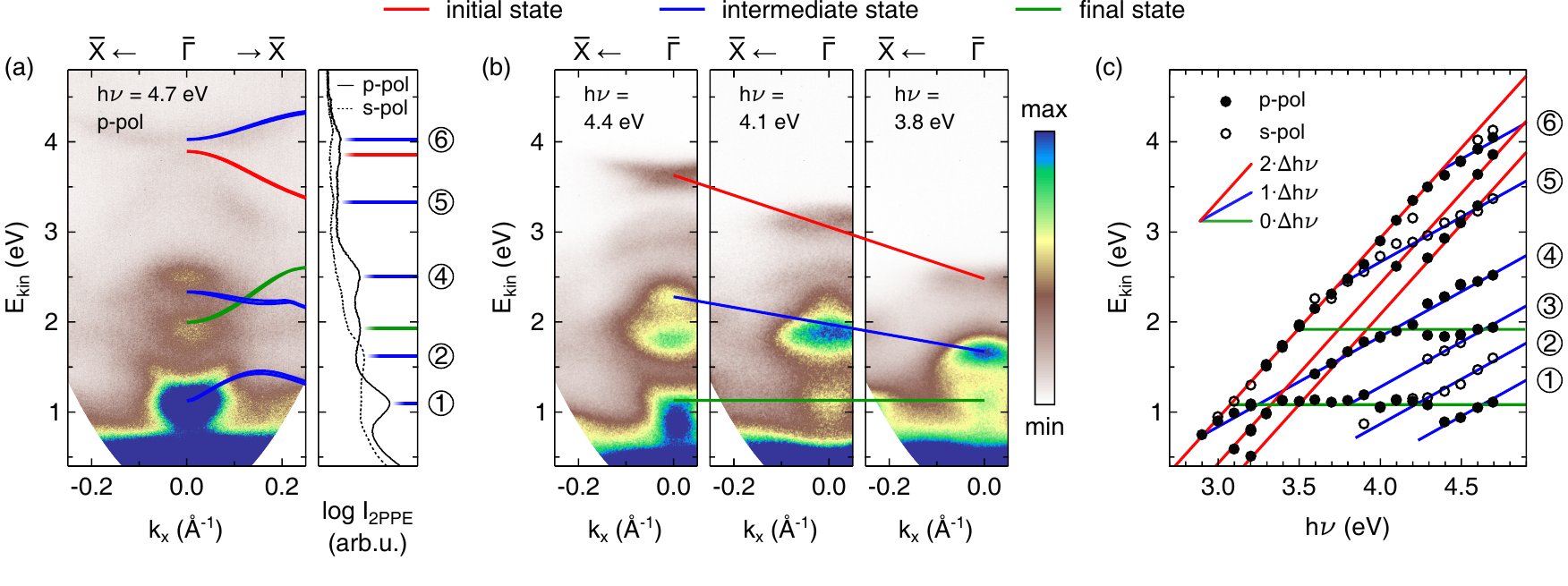}
		\caption{Monochromatic 2PPES data of \textit{Td}-WTe$_2$. (a) 2PPES intensity maps (kinetic energy $E_{\text{kin}}$ vs. surface parallel momentum $k_{\text{x}}$) along $\overbar{\Gamma}$-$\overbar{\text{X}}$ recorded with p-polarized light at $h\nu=\SI{4.7}{\electronvolt}$. EDCs derived from this intensity map (solid line) as well as from an intensity map recorded with s-polarized light (dashed line) are shown to the right. 2PPES intensities were integrated over a momentum window of $\SI{\pm 0.025}{\per\angstrom}$ around $\overbar{\Gamma}$. The 2PPES data are overlaid with results of the band structure calculations along $\Gamma$-$\text{X}$. Only bands are shown that were clearly identified in the p-polarized experimental data. Red lines indicate initial states, blue lines indicate intermediate states, green lines indicate final states. The color-coded markers in the EDCs indicate the corresponding spectral peaks. (b) Comparison of 2PPES intensity maps along $\overbar{\Gamma}$-$\overbar{\text{X}}$ recorded at different photon energies. Data were recorded with p-polarized light. The color-coded lines indicate the kinetic energy shifts expected for initial states (red line), intermediate states (blue line) and final states (green line). (c) Peak kinetic energies at $\overbar{\Gamma}$ as a function of $h\nu$ deduced from Gaussian multi-peak fits of p-polarized and s-polarized monochromatic 2PPES data along $\overbar{\Gamma}$-$\overbar{\text{X}}$. The numbering of the bands allows for better comparison with the band structure data in Fig. \ref{fig:figure_01} and the  experimental results shown in Fig. \ref{fig:figure_06}.} 
		\label{fig:figure_05}
	\end{figure*}
	
	Figure \ref{fig:figure_04}(a) compares bichromatic 2PPES data along $\overbar{\Gamma}$-$\overbar{\text{X}}$ recorded at temporal overlap of a p-polarized \SI{1.7}{\electronvolt}-pump pulse and a s-polarized UV-probe pulse (temporal delay $\Delta t = \SI{0}{\femto\second}$) with data recorded well before the excitation, i.e., at $\Delta t = \SI{-1}{\pico\second}$. The excitation by the intense pump pulse transiently populates states above the Fermi energy $E_\text{F}$ and allows, therefore, addressing part of the unoccupied electronic structure. The transient spectral signal in the excited state regime, which is barely visible in the raw data, becomes substantially enhanced in a difference intensity representation as shown in Fig. \ref{fig:figure_04}(b). The graph covers the excited state regime up to binding energies of $E-E_\text{F}=\SI{1.6}{\electronvolt}$ and was compiled by subtracting 2PPES intensity maps recorded at $\Delta t = \SI{-1}{\pico\second}$ from 2PPES intensity maps recorded at $\Delta t = \SI{0}{\femto\second}$. In the data, we resolve a hole-like band crossing $E_\text{F}$ at $k_\text{x} \approx \pm\SI{0.2}{\per\angstrom}$ and an electron-like band with an energy minimum at $E-E_\text{F} \approx \SI{0.4}{\electronvolt}$. The hole-like band agrees well with the results of the band structure calculations [see solid white lines in Fig. \ref{fig:figure_04}(b)] and was observed in a time-resolved ARPES study of \textit{Td}-WTe$_2$ before \cite{Caputo2018}. However, the band structure calculations do not yield any indication for an electron-like band in the relevant energy-momentum range. Remarkably, a change in the photon energy of the pump pulse by $\Delta h\nu$ results in a corresponding shift of the kinetic energy at which the electron-like parabola appears in the spectra [see peak in the difference energy distribution curves (EDCs) at $\overbar{\Gamma}$ indicated by the blue markers in Fig. \ref{fig:figure_04}(c)]. This observation implies that for the detection of the electron-like band the role of pump and probe pulses are inverted and that the final photoemission process determining the kinetic energy of the photoelectrons is now due to the NIR / VIS pulse [see Fig. \ref{fig:figure_04}(d)]. The reanalysis of the data under consideration of the inversion of pump and probe process results in a corrected value of $E-E_\text{F} \approx \SI{4.6}{\electronvolt}$ for the energy minimum of the electron-like band. The dashed red line in Fig. \ref{fig:figure_04}(b) shows for comparison the results of the band structure calculations along $\Gamma$-$\text{X}$ in the corresponding energy range, which match the experimental data reasonably well.
	
	\subsection{Monochromatic 2PPES results}
	
	Figure \ref{fig:figure_05}(a) shows monochromatic 2PPES intensity maps along $\overbar{\Gamma}$-$\overbar{\text{X}}$ recorded with p-polarized light at $h\nu=\SI{4.7}{\electronvolt}$. 
	To the right, the graph additionally includes EDCs at $\overbar{\Gamma}$ extracted from the 2PPES intensity map shown to the left as well as from data recorded with s-polarized light. We observe a multitude of bands, which vary in their intensity and relative peak position as the photon energy is changed [see Fig. \ref{fig:figure_05}(b)].
	
	Spectral peaks of monochromatic 2PPES data can be due to initial states with binding energies  $E<E_\text{F}$, intermediate states with $E_\text{F}<E<E_\text{V}$, or final states with $E>E_\text{V}$. An assignment is possible by analyzing the shift in the peak kinetic energy as a function of photon energy $h\nu$ \cite{Ueba2007}. For monochromatic 2PPES data, the observation of an energy shift by $2\cdot \Delta h \nu$ ($\Delta h \nu$) is indicative for an initial state (intermediate state). Here, $\Delta h \nu$ denotes the change in the photon energy. In contrast, final states yield no energy shift in the peak kinetic energy at all. Characteristic kinetic energy shifts are exemplarily indicated by the color-coded straight lines in Fig. \ref{fig:figure_05}(b) for selected spectral features that are due to initial, intermediate, and final states, respectively.
	For the evaluation of the 2PPES data, we performed Gaussian multi-peak fits to the EDCs around $\overbar{\Gamma}$. Figure \ref{fig:figure_05}(c) summarizes peak kinetic energies as a function of $h\nu$ as deduced from EDCs recorded with p-polarized and s-polarized laser light along $\overbar{\Gamma}$-$\overbar{\text{X}}$.  
	The different slopes associated with initial, intermediate, and final states are well resolved in the data and are color-coded correspondingly. The discrimination of the state character allows for an evaluation of the state binding energies and finally an assignment to the results of the band structure calculation, partly under consideration of the band dispersion along $\Gamma$-$\text{X}$. Results of this analysis are included in Fig. \ref{fig:figure_05}(a), with an initial state shown as red line, intermediate states shown as blue lines, and a final state shown as green line. Note that only calculated bands are included which could uniquely be identified in the p-polarized experimental data. 

\begin{figure}[t]
		\centering
		\includegraphics[width=1\linewidth]{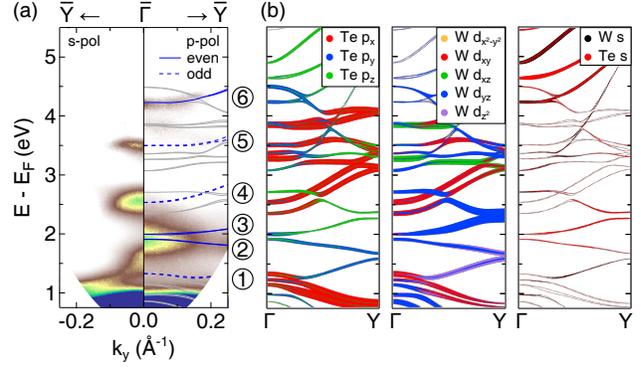}
		\caption{Symmetry properties of intermediate states. (a) Comparison of monochromatic 2PPES intensity maps (binding energy $E-E_{\text{F}}$ vs. surface parallel momentum $k_\text{y}$) recorded at $h\nu=\SI{4.5}{\electronvolt}$ along $\overbar{\Gamma}$-$\overbar{\text{Y}}$ with s- and p-polarized light. The experimental data are overlaid with the results of band structure calculations in the corresponding intermediate state energy range. Bands that could be assigned based on the kinetic energy analysis of the 2PPES data are indicated in blue, with solid lines indicating bands of even symmetry and dashed lines indicating bands of odd symmetry with respect to the mirror plane of the \textit{Td} structure. The numbering of the bands allows for better comparison with the band structure data in Fig. \ref{fig:figure_01} and the  experimental results shown in Fig. \ref{fig:figure_05}. (b) Calculated Tellurium and Tungsten contributions to the bands along $\Gamma$-$\text{Y}$ in the energy range covered by the monochromatic 2PPES data. The size of the data points indicates the respective orbital contribution to the individual bands. In Fig. S2 of the supplemental material, the orbital-resolved data are shown separately.}
		\label{fig:figure_06}
	\end{figure}

	By analogy with the ARPES data, the polarization dependence of the monochromatic 2PPES signal in normal emission along $\overbar{\Gamma}$-$\overbar{\text{Y}}$ contains information about the intermediate state symmetry with respect to the mirror plane of the \textit{Td} structure of WTe$_2$. Figure \ref{fig:figure_06}(a) compares monochromatic 2PPES data recorded along $\overbar{\Gamma}$-$\overbar{\text{Y}}$ at a photon energy of $h\nu = \SI{4.5}{\electronvolt}$ with s- and p-polarization, respectively. We selected this data set for the symmetry analysis as for a photon energy of $h\nu = \SI{4.5}{\electronvolt}$ all six bands associated with bands in the intermediate state energy range are visible in the spectra [see Fig. \ref{fig:figure_05}(c)]. The intensity map is overlaid with the intermediate state band structure along $\Gamma$-$\text{Y}$ with the bands highlighted that could be assigned based on the kinetic energy analysis of the 2PPES data. Figure \ref{fig:figure_06}(b) shows in comparison the calculated orbital-resolved band structure data. The lowest energy band at $E-E_{F}\approx \SI{1.3}{\electronvolt}$ is visible with s-polarized light and lies at $\overbar{\Gamma}$ close to a final state resonance. Still, at finite momentum we can clearly identify the downward dispersing character of the band as predicted by theory. Near $\Gamma$, the band is predominantly formed from $\text{Te}$-$p_x$ and $\text{W}$-$d_{xy}$ orbitals and exhibits therefore an odd symmetry with respect to the mirror plane in accordance with the observation with s-polarized light. The same orbital character is also found for the band visible in the s-polarized data at $E-E_{F}\approx \SI{2.5}{\electronvolt}$. Finally, the odd symmetry of the band seen at $E-E_{F}\approx \SI{3.5}{\electronvolt}$ arises from a $\text{Te}$-$p_x$ orbital character. In the p-polarized data set, we identify two unoccupied bands at $E-E_{F}\approx \SI{2}{\electronvolt}$ and a further unoccupied band at $E-E_{F}\approx \SI{4.2}{\electronvolt}$. The orbital character of the two low energy bands is predominantly $\text{Te}$-$p_y$ and $\text{W}$-$d_{yz}$ (with an increasing $\text{W}$-$d_{z^2}$ contribution at high momentum for the lower band). The even symmetry of the high energy band mainly arises from $\text{Te}$-$p_y$ orbitals and some contributions from $\text{Te}$-$s$ and $\text{W}$-$s$ orbitals. Overall, the accordance between the polarization dependence of the 2PPES spectra along $\overbar{\Gamma}$-$\overbar{\text{Y}}$ and the orbital character predicted by the band structure calculations confirms the band assignment that we made based on the kinetic energy and dispersion analysis of the experimental data.                    
	
	Finally, two final state resonances can be identified in the monochromatic 2PPES data [see Fig. \ref{fig:figure_05}(c)]. The observed kinetic energies correspond to binding energies of $E-E_\text{F}\approx \SI{6.0}{\electronvolt}$ and $E-E_\text{F}\approx \SI{6.9}{\electronvolt}$. A convincing assignment of the latter state to results of the band structure calculations is possible [cf. Fig. \ref{fig:figure_01}(c)] and is correspondingly included in the 2PPES spectra shown in Fig. \ref{fig:figure_05}(a). However, the band structure calculations do not yield any indication for a band near $E-E_\text{F}\approx \SI{6.0}{\electronvolt}$. The resonance is instead located within a surface projected band gap at $\overbar{\Gamma}$ [see green shaded area in Fig. \ref{fig:figure_01}(c)] indicating that we probe here a surface located resonance in the final state energy range.

	\section{\label{sec:level4}Conclusion}
	In combining laser-based ARPES and 2PPES, we studied in this work the valence electronic structure of the Weyl semimetal candidate \textit{Td}-WTe$_2$ along two high symmetry directions of the crystalline structure. The low photon energy provided by the laser system considerably limits the accessible momentum range to values close to $\overbar{\Gamma}$, however, the bichromatic and monochromatic 2PPES data enabled us to access the unoccupied electronic structure up to binding energies of $E-E_{\text{F}}\approx \SI{5}{\electronvolt}$. Furthermore, the high flexibility in tuning the polarization state of the laser light enabled us to experimentally gain information on initial and intermediate state symmetries under consideration of the dipole selection rules in photoemission. Based on these results and in comparison with band structure calculations, it was possible to identify the predominant orbital contributions to the ARPES and 2PPES signal for photoemission within the mirror plane of the \textit{Td} structure of WTe$_2$.

	\section{\label{sec:level5}Acknowledgments}
	This work was supported by the German Research Foundation (DFG) through projects INST 257/419-1 and INST 257/442-1
	
	\section*{References}
	\bibliography{MyCollection}
	
\end{document}


	
	\title{Supplemental Material for:\\
			"A combined ARPES and two-photon photoemission study of \textit{Td}-\ce{WTe2}"}

	\author{Petra Hein}
	\email[]{hein@physik.uni-kiel.de}
	\affiliation{Institute of Experimental and Applied Physics, University of Kiel, Leibnizstr. 19, D-24118 Kiel, Germany}
	
	\author{Stephan Jauernik}
	\affiliation{Institute of Experimental and Applied Physics, University of Kiel, Leibnizstr. 19, D-24118 Kiel, Germany}
	
	\author{Hermann Erk}
	\affiliation{Institute of Experimental and Applied Physics, University of Kiel, Leibnizstr. 19, D-24118 Kiel, Germany}
	
	\author{Lexian Yang}
	\affiliation{State Key Laboratory of Low Dimensional Quantum Physics, Collaborative Innovation Center of Quantum Matter and Department of Physics, Tsinghua University, Beijing 100084, China}
	
	\author{Yanpeng Qi}
	\affiliation{School of Physical Science and Technology, ShanghaiTech University, Shanghai 201210, China}
	\affiliation{Max Planck Institute for Chemical Physics of Solids, N\"othnitzer Str. 40, D-01187 Dresden, Germany}
	
	\author{Yan Sun}
	\affiliation{Max Planck Institute for Chemical Physics of Solids, N\"othnitzer Str. 40, D-01187 Dresden, Germany}
	
	\author{Claudia Felser}
	\affiliation{Max Planck Institute for Chemical Physics of Solids, N\"othnitzer Str. 40, D-01187 Dresden, Germany}
	
	\author{Michael Bauer}
	\affiliation{Institute of Experimental and Applied Physics, University of Kiel, Leibnizstr. 19, D-24118 Kiel, Germany}
	%
	%
	%
	
	\date{\today}

	\maketitle
	
\newpage
	
\section{\label{sec:level1}Band structure of \textit{Td}-\ce{WTe2}}

\begin{figure}[h!]
	\centering
	\includegraphics[width=1\linewidth]{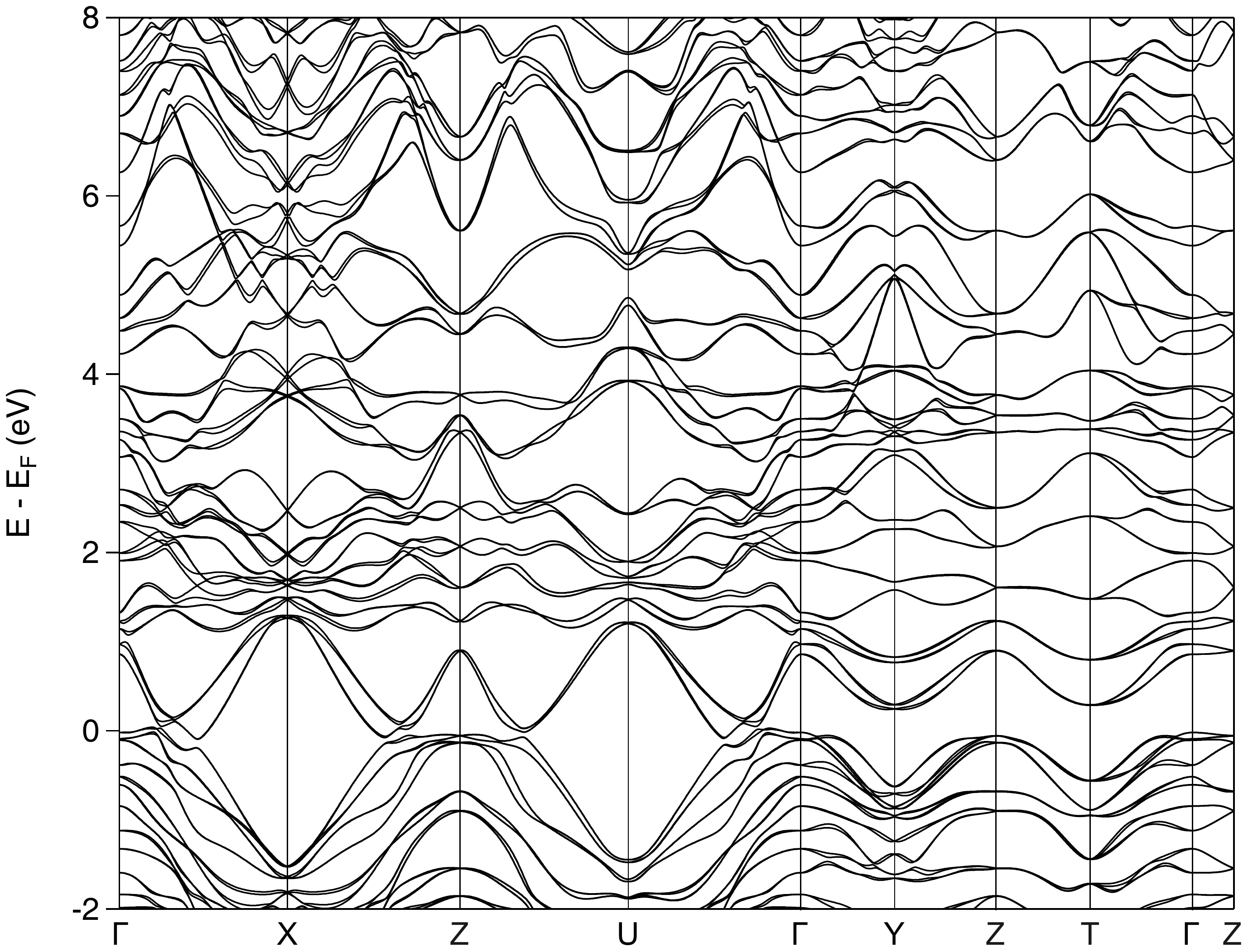}
	\caption{Calculated bulk band structure of \textit{Td}-WTe$_2$ along high symmetry directions of the Brillouin zone.}
	\label{fig:Suppl_figure_01}
\end{figure}

\newpage
\section{\label{sec:level2}Orbital-resolved band structures}

\begin{figure}[h!]
	\centering
	\includegraphics[width=0.9\linewidth]{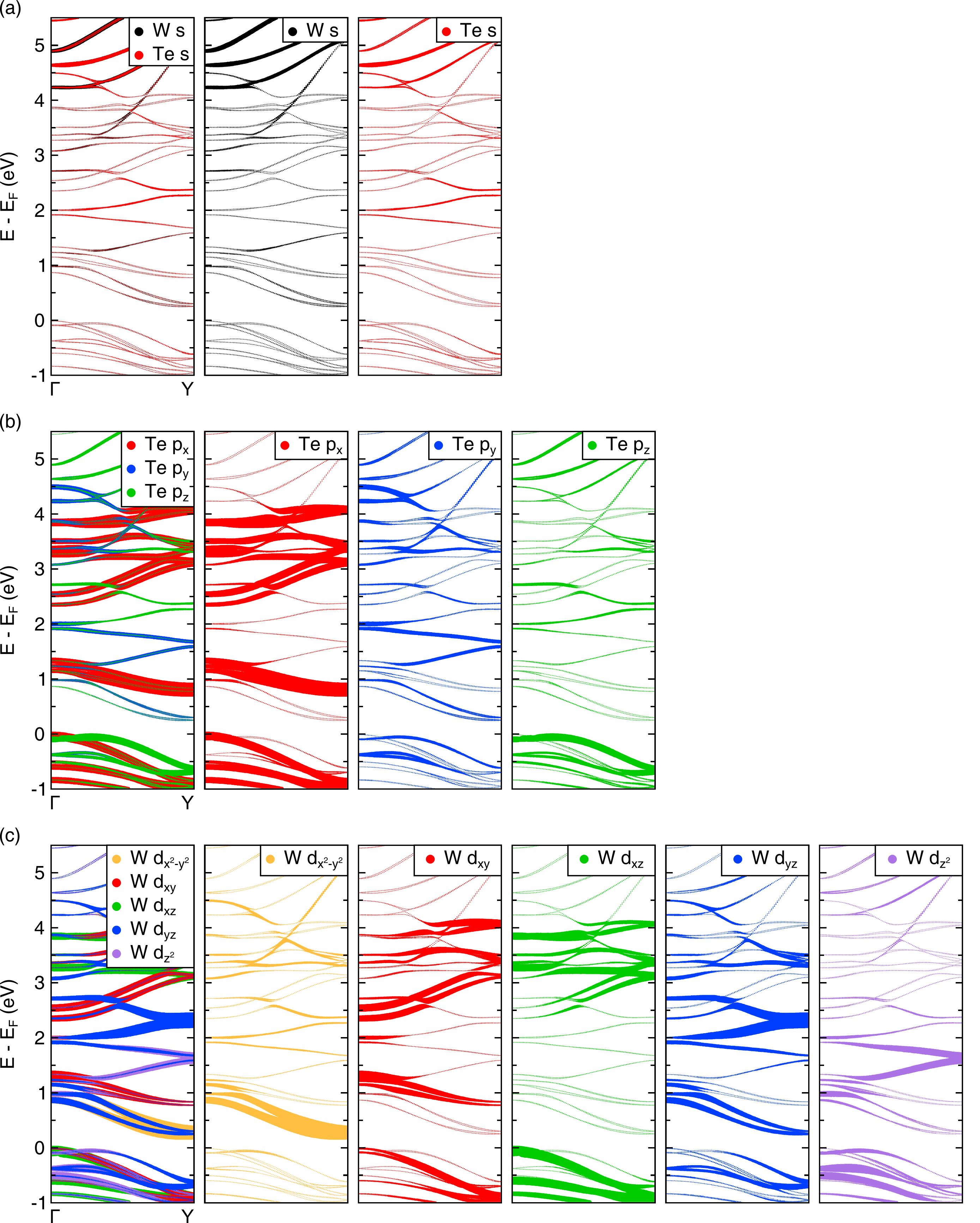}
	\caption{Calculated orbital-resolved band structures along $\Gamma$-$\text{Y}$ for the initial and intermediate state energy range covered by the ARPES and 2PPES data. The size of the data points indicates the respective orbital contribution to the individual bands.}
	\label{fig:Suppl_figure_02}
\end{figure}


